\documentclass[10pt,american,english]{IEEEtran}
\usepackage[LGR,T1]{fontenc}
\usepackage{textcomp}
\usepackage[utf8]{inputenc}
\usepackage{float}
\usepackage{amsmath}
\usepackage{amssymb}
\usepackage{graphicx}
\usepackage{setspace}
\usepackage{wasysym}
\usepackage{flushend}

\makeatletter

\DeclareRobustCommand{\greektext}{%
  \fontencoding{LGR}\selectfont\def\encodingdefault{LGR}}
\DeclareRobustCommand{\textgreek}[1]{\leavevmode{\greektext #1}}

\newcommand{\lyxmathsym}[1]{\ifmmode\begingroup\def\b@ld{bold}
  \text{\ifx\math@version\b@ld\bfseries\fi#1}\endgroup\else#1\fi}

\floatstyle{ruled}
\newfloat{algorithm}{tbp}{loa}
\providecommand{\algorithmname}{Algorithm}
\floatname{algorithm}{\protect\algorithmname}

\usepackage{algorithm}
\usepackage{algorithmic}

\makeatother

\usepackage{babel}
\addto\captionsenglish{\renewcommand{\algorithmname}{Algorithm}}

\begin{document}
\title{Secure Communication of UAV-mounted STAR-RIS under
Phase Shift Errors}
\author{Aseel Qsibat, Habiba Akhleifa, Abdelhamid Salem, Khaled Rabie, \textit{Senior Member, IEEE}, Xingwang Li, \textit{Senior Member, IEEE}, Thokozani Shongwe,  Mohamad A. Alawad, and Yazeed Alkhrijah

\thanks{Corresponding authors: A. Salem and Mohamad A. Alawad.}

\thanks{Aseel Qsibat and Habiba Akhleifa are with the Department of Electronic and Electrical Engineering, Benghazi University, Benghazi 1308, Libya, (e-mails: aseelqsibat@gmail.com, habiba.akhlifa@gmail.com).}

\thanks{A. Salem is with the Department of Electronic and Electrical Engineering, University College London, London, UK, (e-mail: a.salem@ucl.ac.uk). A. Salem is also affiliated with the Department of Electronic and Electrical Engineering, Benghazi University, Benghazi 1308, Libya.}

\thanks{Khaled Rabie is with the Department of Computer Engineering, King Fahd University of Petroleum and Minerals (KFUPM), Dhahran, Saudi Arabia (email: k.rabie@kfupm.edu.sa).}

\thanks{X. Li is with the School of Physics and Electronic Information Engineering, Henan Polytechnic University, Jiaozuo 454003, China, and also with National Mobile Communications Research Laboratory, Southeast University, Nanjing 210096, China (e-mail: lixingwangbupt@gmail.com).}

\thanks{Thokozani Shongwe is with the Department of Electrical and Electronic Engineering Technology, University of Johannesburg, South Africa.}

\thanks{Mohamad A. Alawad is with the Department of Electrical Engineering, Imam Mohammad Ibn Saud Islamic University (IMSIU), Riyadh, Saudi Arabia, (email: maawaad@imamu.edu.sa).}

\thanks{Yazeed Alkhrijah is with the Department of Electrical Engineering, College of Engineering, Imam Mohammad Ibn Saud Islamic University (IMSIU), Riyadh, Saudi Arabia, (email: ymalkhrijah@imamu.edu.sa).}

}
\maketitle
\begin{abstract}
This paper investigates the secure communication capabilities of a non-orthogonal multiple access (NOMA) network supported by a STAR-RIS (simultaneously transmitting and reflecting reconfigurable intelligent surface) deployed on an unmanned aerial vehicle (UAV), in the presence of passive eavesdroppers. The STAR-RIS facilitates concurrent signal reflection and transmission, allowing multiple legitimate users—grouped via NOMA—to be served efficiently, thereby improving spectral utilization. Each user contends with an associated eavesdropper, creating a stringent security scenario. Under Nakagami fading conditions and accounting for phase shift inaccuracies in the STAR-RIS, closed-form expressions for the ergodic secrecy rates of users in both transmission and reflection paths are derived. An optimization framework is then developed to jointly adjust the UAV’s positioning and the STAR-RIS power splitting coefficient, aiming to maximize the system’s secrecy rate. The proposed approach enhances secure transmission in STAR-RIS-NOMA configurations under realistic hardware constraints and offers valuable guidance for the design of future 6G wireless networks.
\end{abstract}


\section{INTRODUCTION}

\IEEEPARstart{W}{ith} the global rollout of fifth-generation (5G)
wireless networks, researchers are now focusing on the development
of next-generation communication systems, namely, sixth-generation
(6G) networks. These advancements aim to to meet increasingly stringent demands, including enhanced spectral efficiency, improved energy efficiency (EE), ultra-low latency, and expanded coverage \cite{re1}
 all to ensure the desired quality of service (QoS).

Among others, the advent of reconfigurable intelligent surfaces (RIS)
has introduced a revolutionary technology for augmenting spectral
and energy efficiency \cite{re2}.  RIS is made up of a flat array of inexpensive passive components, where each element can control electromagnetic (EM) waves by adjusting their phase shifts using built-in PIN diodes  \cite{re3}. By dynamically tuning these phase shifts via
an intelligent controller, the RIS can enhance signal strength at
intended receivers while suppressing interference or signals directed
toward eavesdroppers. Unlike traditional techniques---such as artificial
noise injection \cite{re4} or multi-antenna beamforming \cite{re5}---RIS
operates passively without requiring additional RF chains, making
it both energy-efficient and cost-effective. Moreover, RIS can be
seamlessly integrated into existing infrastructure by mounting it
on surfaces like building facades, windows, or even clothing, offering
flexible deployment options \cite{re10}.

However, conventional RIS technology only reflects incoming signals,
restricting communication to one side of the surface (the reflection
plane) \cite{re11}. Early investigations into the practical realization of metasurfaces capable of both transmission and reflection were presented in \cite{re14}. Building on these findings, the adoption of STAR-RIS in wireless communication has shown practical potential, though it introduces a complex and adversarial operating environment. In contrast to conventional RIS technology, STAR-RIS is capable of independently adjusting both the transmission and reflection coefficients at each element. This capability allows a single transmitter to simultaneously serve users located on both sides of the RIS, effectively enabling 360-degree coverage and overcoming the directional limitations of traditional RIS systems \cite{re11}. Numerous studies have explored the performance improvements offered by RIS in various communication contexts. For example, \cite{re014} developed a low-complexity strategy for managing the trade-off between sum-rate and energy usage in a RIS-enhanced NOMA system, employing the successive convex approximation (SCA) technique to alternately handle beamforming and phase shift optimization. In a separate study, the authors in 
\cite{re017}  focused on improving the long-term energy efficiency (EE) of STAR-RIS-enabled CoMP networks by jointly optimizing active and passive beamforming, incorporating fractional programming and deep reinforcement learning (DRL) to determine an adaptive, near-optimal solution. Moreover, both perfect and imperfect RIS conditions were explored in \cite{re016}, within CoMP scenarios, where optimal reflection parameters were determined using a dual Lagrangian framework. In \cite{re015}, analyzed the performance enhancement of RIS in semi-grant-free NOMA systems, proposing customized algorithms tailored to various RIS configurations. Furthermore, the authors in \cite{re018} addressed the sum-rate maximization problem in a multi-UAV NOMA network by jointly optimizing UAV deployment, transmit power, RIS reflection matrix, and NOMA decoding order. To efficiently solve this complex problem, a block coordinate descent (BCD)-based iterative algorithm was developed. Additionally,  \cite{re019} proposed an innovative distributionally robust reinforcement learning (DRRL) approach, which simultaneously optimizes the UAV's flight path, the active beamforming at the UAV, and the passive beamforming at the STAR-RIS. This integrated design aims to improve the performance of UAV communication systems supported by STAR-RIS.
Meanwhile, UAVs have emerged as crucial enablers for future wireless
networks due to their cost-effectiveness, rapid deployment, and mobility
\cite{re29}, serving as aerial base stations, relays, or wireless
chargers to enhance coverage and capacity \cite{re30}. Recent advances
combine UAVs with RIS, creating mobile RIS platforms that offer dynamic
beamforming capabilities \cite{re31}. This integration appears in
two primary configurations: ground-based RIS assisting UAV-ground
communications \cite{re35}, \cite{re36} and aerial RIS mounted on
UAVs for omnidirectional coverage \cite{re30}. While UAV-mounted
RIS provides deployment flexibility and improved LoS connectivity
\cite{re32}, it introduces complex co-design challenges for RIS configuration
and UAV positioning. These hybrid systems present new opportunities
to enhance both coverage and security in next-generation networks,
though practical implementation hurdles remain regarding real-time
optimization of RIS parameters and UAV trajectory planning in dynamic
environments. At this point, the role of physical layer security (PLS)
becomes crucial in addressing these challenges by leveraging the inherent
characteristics of the wireless medium to provide low-complexity and
real-time adaptable security solutions.

The PLS enhances data confidentiality by exploiting the inherent randomness of wireless channels, complementing traditional cryptographic methods \cite{re01}. Integrating RIS into PLS frameworks introduces several key benefits. In particular, RIS enables passive beamforming through its passive elements, allowing precise control over signal propagation and resulting in significant improvements in secrecy performance \cite{re02}. Additionally, RIS can provide signal coverage in environments where direct links are obstructed \cite{re03},
making it especially suitable for deployment near legitimate users in blocked or non-line-of-sight (NLoS) scenarios. Its compatibility with existing infrastructure further reduces deployment costs and complexity \cite{re04}. In this context, 
\cite{re05} analyzed the secrecy rate maximization of an RIS-assisted multi-antenna system under constraints of transmission power and RIS phase shifts. Building on this, \cite{re06} proposed a power-efficient secure transmission design. The work in \cite{re07},
investigated the joint use of RIS and artificial noise (AN) to further enhance PLS. Moreover, \cite{re08} explored the application of RIS in device-to-device and relay-based systems, demonstrating its effectiveness in improving secrecy performance across diverse communication scenarios.

Several studies have examined the potential of RIS-assisted secure communication systems, leveraging a combination of active transmit and passive reflective beamforming along with various optimization techniques to enhance the secrecy rate. For instance, in \cite{re001}, the authors investigated the secrecy performance of RIS-assisted mmWave/THz communication systems by jointly designing transmit beamforming and discrete phase shifts to maximize the secrecy rate. In \cite{re002}, the impact of hardware impairments was considered in the context of RIS-aided PLS networks with multi-antenna transmitters. Expanding on this, \cite{re003} analyzed a secure communication scenario involving multiple multi-antenna eavesdroppers and a NLoS channel between the access point and the legitimate user. Collectively, these works highlight the effectiveness of RIS in strengthening physical layer security while maintaining energy efficiency. Furthermore, they suggest that uniformly distributing the reflecting elements across multiple RIS units is a promising strategy for boosting PLS and harnessing macro-diversity gains.

The deployment of flying RIS is typically constrained to ensure that both the transmitter and receiver remain on the same side of the surface, which inherently limits the achievable system performance. To address this limitation, STAR-RIS has been introduced to enable full-space coverage. STAR-RIS achieves this by concurrently reflecting and transmitting incident signals toward both the same-side and opposite-side half-spaces relative to the source node, as demonstrated in  \cite{re19}, \cite{re37}. Furthermore, STAR-RIS and NOMA exhibit a mutually beneficial relationship—STAR-RIS can significantly expand the coverage area of NOMA systems, while NOMA can enhance the overall spectral efficiency and user access capabilities of STAR-RIS-assisted networks \cite{re26}, \cite{re15}, \cite{re27}. However,
despite the transformative potential of STAR-RIS, especially when
deployed on UAV platforms, the PLS performance of such systems remains
largely unexplored. This critical research gap is further exacerbated
by real-world challenges---most notably, the phase distortion caused
by UAV vibrations and aerodynamic turbulence, which introduces non-negligible
phase estimation errors that degrade beamforming accuracy and secrecy
performance.

Motivated by this, in this work, we conduct a detailed performance evaluation of a UAV-enabled communication system incorporating a STAR-RIS. By harnessing the UAV's mobility alongside the STAR-RIS’s full-space signal manipulation capability, we aim to boost the secrecy performance of NOMA networks—particularly in scenarios affected by phase estimation errors resulting from UAV instability and airflow fluctuations. We also explore optimal strategies for power allocation and the placement of both the UAV and STAR-RIS in environments involving multiple legitimate users and eavesdroppers.

The main contributions of this paper are summarized as follows:
\begin{itemize}
\item We investigate a UAV-mounted STAR-RIS-assisted downlink NOMA communication
system in the presence of multiple eavesdroppers. To fully exploit
the 360-degree transmission and reflection capability of the STAR-RIS,
a user pairing strategy is proposed to serve users located in both
the reflection and transmission regions via NOMA. Under this framework,
we derive closed-form expressions for the ergodic secrecy rates of
legitimate users in both regions.
\item To realistically account for channel impairments caused by UAV mobility,
we model the phase estimation errors using a von Mises distribution,
which captures practical distortions such as jittering and airflow.
Leveraging this model, we analytically derive ergodic secrecy rate
expressions that are computationally efficient and provide results
consistent with Monte Carlo simulations.
\item To enhance the PLS performance, we formulate a joint optimization
problem to maximize the Weighted Sum Secrecy Rate (WSSR) by simultaneously
optimizing the UAV’s 3D hovering position and transmit power allocation
among NOMA users.
\item Due to the non-convex nature of the formulated problem, we decompose
this problem into two subproblems: one for UAV positioning and the
other for power allocation. An iterative alternating optimization
algorithm is proposed to solve the problem efficiently until convergence.
\item Specifically, the 3D UAV position is optimized using a linear grid-based
search over a discretized space to find the best hovering location,
while the power allocation is optimized using a global search algorithm,
which is well-suited for handling the highly non-convex and continuous
nature of the power control problem in NOMA systems with secrecy constraints.
\end{itemize}
\noindent The remainder of this paper is organized as follows. Section II outlines the system model and formulates the joint optimization framework for a UAV-assisted STAR-RIS downlink NOMA system under a fixed deployment scenario. Section III describes the proposed algorithms developed to address this optimization task. Section V presents simulation results that validate the efficiency of the proposed solutions and demonstrate the benefits of integrating STAR-RIS with UAVs. Lastly, Section VI provides concluding remarks.

\section{SYSTEM MODEL}

\begin{figure}[tp]
\centering{}\includegraphics[scale=0.35]{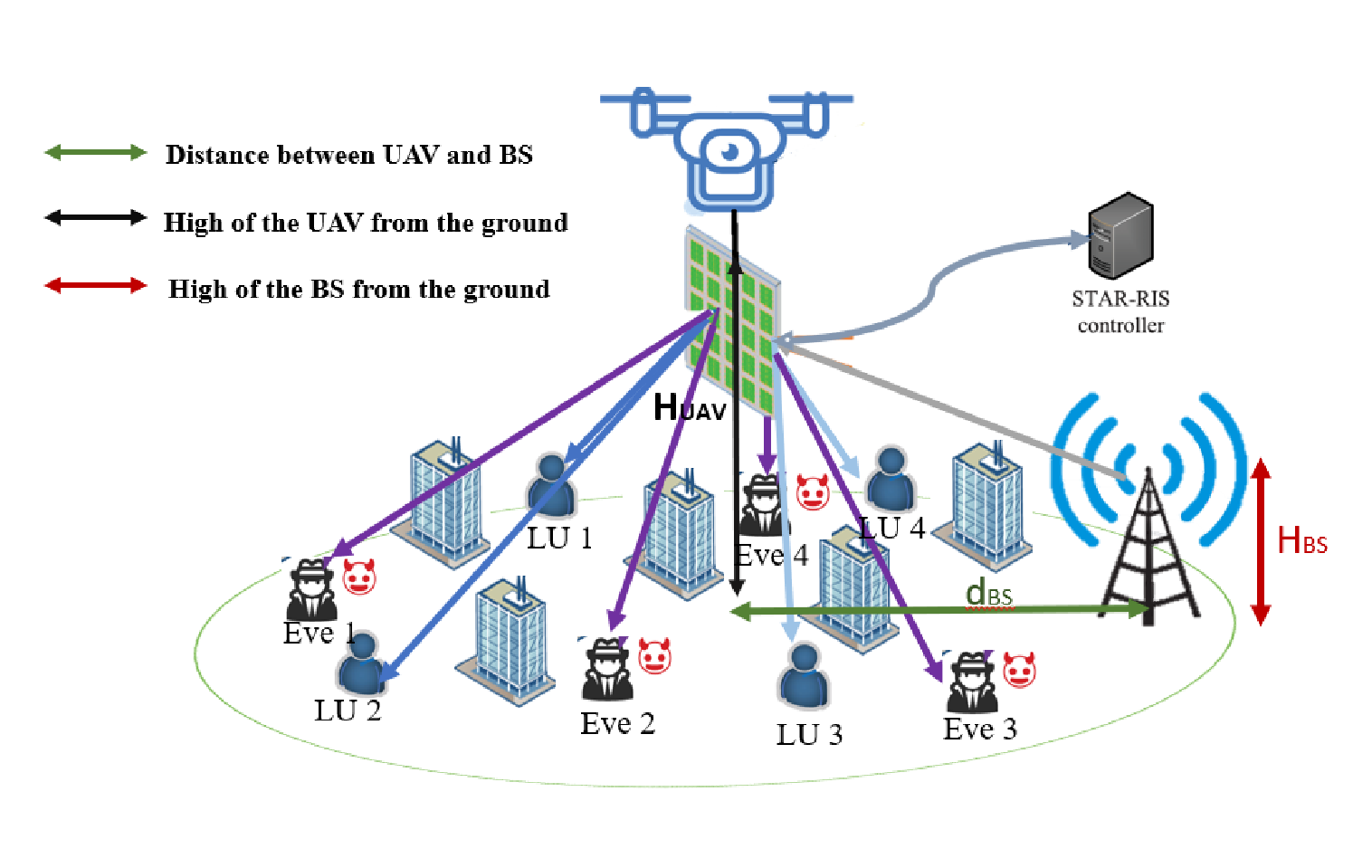}

\caption{UAV-mounted STAR-RIS-assisted Secure Downlink NOMA Communications
System}
\end{figure}

As illustrated in Fig. 1, we consider a secure downlink transmission
scenario in a NOMA system assisted by a UAV-mounted STAR-RIS. A ground
base station (BS) communicates with multiple single-antenna legitimate
users (LUs), which are divided into two distinct groups: the legitimate users are positioned within both the reflection and transmission regions of the STAR-RIS. Considering the presence of eavesdroppers (Eves) distributed across both sides of the STAR-RIS, it is assumed that the direct communication links between the base station (BS) and the users are obstructed by environmental barriers. This assumption reflects a widely adopted worst-case scenario in physical layer security (PLS) analyses. \cite{re26}.

Let the number of legitimate users be denoted by $K=K_{r}+K_{t}$,
where $K_{r}$ and $K_{t}$ represent the number of users in the reflection
and transmission regions, respectively. Similarly, let \ensuremath{E}
be the number of single-antenna eavesdroppers and \ensuremath{M} be
the number of elements on the STAR-RIS. The STAR-RIS, mounted on a
UAV, is capable of adjusting its position and altitude dynamically,
allowing for more adaptable line-of-sight (LoS) connectivity with ground users while improving the system’s spatial responsiveness.
\cite{re43}.

We adopt a NOMA scheme in which each time-frequency resource block
is shared by a user pair---one from the reflection region and one
from the transmission region. Power allocation is asymmetric: the
reflect-region user, typically closer to the UAV, is allocated less
power, while the transmit-region user, being further, receives more
power but is subject to stronger interference. This pairing strategy
aligns with established NOMA protocols, which leverage successive
interference cancellation (SIC) for decoding \cite{re41}, that can
be viewed as the worst-case scenario\footnote{Note that the assumption of single-antenna eavesdroppers represents a fundamental baseline threat model, while scenarios involving multi-antenna eavesdroppers will be studied in our future works.}. In this paper, a 3D Cartesian coordinate system is adopted. 
The horizontal locations of LU $k\in\mathcal{K}$ and Eve $e\in\mathcal{E}$
are denoted by $C_{k}$ = {[}$x_{k}$, $y_{k}$, $z_{k}${]} and $C_{e}$
= {[}$x_{e}$, $y_{e}$, $z_{e}${]}, respectively. Moreover, the
location of the UAV is denoted by $C_{u}$ = {[}$x_{u}$, $y_{u}$,
$z_{u}${]}.

All wireless channels are modeled using Nakagami-\ensuremath{m} fading,
which provides a flexible and generalized fading framework appropriate
for various propagation conditions. To account for practical impairments
in STAR-RIS implementation, phase errors introduced by the STAR-RIS
elements are modeled as von Mises-distributed random variables, capturing
the imperfections in phase control due to hardware limitations \cite{re42}.

In the context of the STAR-RIS, the transmission coefficient (TCs) and reflection coefficient (RCs) matrices are denoted as $\Theta_{t}\in\mathbb{C}^{M\times M}$
and $\Theta_{r}\in\mathbb{C}^{M\times M}$, respectively. Correspondingly,
we have $\Theta_{t}=$ diag $\left\{ \sqrt{\zeta_{t}^{1}}e^{-j\theta_{t}^{1}},...,\sqrt{\zeta_{t}^{M}}e^{-j\theta_{t}^{M}}\right\} $
and $\Theta_{r}=$ diag $\left\{ \sqrt{\zeta_{r}^{1}}e^{-j\theta_{t}^{1}},...,\sqrt{\zeta_{r}^{M}}e^{-j\theta_{r}^{M}}\right\} $,
where $\zeta_{q}^{m}\in[0,1]$ and $\Theta_{q}^{m}\in[0,2\pi)$, $q\in[r,t]$
are the amplitude and phase shift of element m, respectively. Similar
to \cite{re52}, the UAV-mounted STAR-RIS communicates with the controller through wireless backhaul links, which are used to configure its transmission and reflection coefficients. Specifically, the amplitudes $\zeta_{t}^{m}$ and
$\zeta_{r}^{m}$ are coupled to satisfy the conditions of $\zeta_{t}^{m}+\zeta_{r}^{m}=1$
due to the law of energy conservation, while the phase shifts $\Theta_{t}^{m}$
$\Theta_{r}^{m}$ can be independently adjusted. Similar to \cite{re27},
it is assumed that the direct links between the legitimate users (LUs) and the base station (BS) are obstructed by environmental obstacles, representing one of the most challenging conditions for conventional communication systems. Furthermore, it is assumed that the UAV has access to the location information of each legitimate user (LU) and eavesdropper (Eve) through the use of on-board optical cameras or synthetic aperture radar. \cite{re53}.

\subsection{Signal Model}

In the considered STAR-RIS-assisted UAV communication system, NOMA
is employed to enhance spectral efficiency by allowing simultaneous
transmission to multiple users over the same frequency band. In this
framework, the BS transmits a superimposed signal composed of different
user messages, each assigned a distinct power level based on their
channel conditions. Specifically, the BS allocates a portion of its
total transmit power $P_{s}$ to serve users in both the reflection
and transmission regions of the STAR-RIS. The transmitted signal is
given by:
\begin{equation}
S=\sqrt{\rho P_{s}}x_{r}+\sqrt{(1-\rho)P_{s}}x_{t},
\end{equation}
where $x_{r}$ and $x_{t}$ present the information signals intended
for users in the reflection and transmission regions, respectively,
using power-domain NOMA $\rho\in[0,1].$

\noindent The cascaded channels from the BS to a user (or eavesdropper)
involve the links from BS to STAR-RIS and STAR-RIS to receiver. $h_{i}$
and $\{g_{k_{q_{i}}}/g_{e_{q_{i}}}\}$ represent the channel from
the BS to the RIS and the channel from the ground RIS to LUs/Eves,
where $q\in[r,t]$. Therefore, after applying SIC, the received signal
at the legitimate users $k_{r}$ at reflect area and $k_{t}$ at transmit
area can be expressed as:

\begin{equation}
y_{k_{r}}=\sqrt{\frac{\rho\zeta P_{s}}{d_{BV}^{\alpha}d_{VU_{r}}^{\alpha}}}\left[\sum_{i=1}^{M}h_{i}e^{-j\theta_{r}^{m}}g_{r_{i}}\right]x_{r}+n_{r},
\end{equation}

\[
y_{k_{t}}=\sqrt{\frac{(1-\rho)(1-\zeta)P_{s}}{d_{BV}^{\alpha}d_{VU_{t}}^{\alpha}}}\left[\sum_{i=1}^{M}h_{i}e^{-j\theta_{t}^{m}}g_{t_{i}}\right]x_{t}
\]

\begin{equation}
+\sqrt{\frac{\rho\zeta P_{s}}{d_{BV}^{\alpha}d_{VE}^{\alpha}}}\left[\sum_{i=1}^{M}h_{i}e^{-j\theta_{t}^{m}}g_{t_{i}}\right]x_{r}+n_{t}.
\end{equation}

Similarly, the received signals at each Eavesdropper in both areas
are given by:

\begin{equation}
y_{e_{r}}=\sqrt{\frac{\rho\zeta P_{s}}{d_{BV}^{\alpha}d_{VE}^{\alpha}}}\left[\sum_{i=1}^{M}h_{i}e^{-j\theta_{r}^{m}}g_{er_{i}}\right]x_{r}+n_{er},
\end{equation}

\[
y_{e_{t}}=\sqrt{\frac{(1-\rho)(1-\zeta)P_{s}}{d_{BV}^{\alpha}d_{VE}^{\alpha}}}\left[\sum_{i=1}^{M}h_{i}e^{-j\theta_{t}^{m}}g_{et_{i}}\right]x_{t}
\]

\begin{equation}
+\sqrt{\frac{\rho\zeta P_{s}}{d_{BV}^{\alpha}d_{VE}^{\alpha}}}\left[\sum_{i=1}^{M}h_{i}e^{-j\theta_{t}^{m}}g_{et_{i}}\right]x_{r}+n_{et}.
\end{equation}

\noindent where $d_{BV}=\sqrt{(x_{b}-x_{u})^{2}+(y_{b}-y_{u})^{2}+(z_{b}-z_{u})^{2}}$
denotes the Euclidean distances between the BS and UAV, $d_{VU_{q}}=\sqrt{(x_{k_{q}}-x_{u})^{2}+(y_{k_{q}}-y_{u})^{2}+z_{u}^{2}}$
denotes the Euclidean distances between the STAR-RIS and LU k, and
$d_{VE_{q}}=\sqrt{(x_{e_{q}}-x_{u})^{2}+(y_{e_{q}}-y_{u})^{2}+z_{u}^{2}}$
denotes the Euclidean distances between the STAR-RIS and Eve. \textgreek{ε}
is the corresponding path loss exponents. The phase shifts \textgreek{θ}$_{r}^{m}$
and \textgreek{θ}$_{t}^{m}$ can be independently adjusted. n$\sim$$\mathcal{CN}(0,N_{0})$
represents the additive white Gaussian noise (AWGN) and $N_{0}$ is
the variance. However, imperfections from jittering and airflow make precise adjustment of the reflecting elements challenging, resulting in practical phase shifts with error $\phi$,
thus:
\begin{equation}
\theta{}_{q}^{m}=\theta_{opt}+\phi.
\end{equation}

\noindent Under the NOMA framework, LUs perform successive interference cancellation (SIC) according to a decoding order. Without loss of generality, this order is assumed to correspond to descending channel gains across the K LUs. Then, the signal-interference-to-noise
ratio (SINR) at the LUs $\{k_{r},k_{t}\}$ using (2) and (3), can be
given by:

\begin{equation}
\gamma_{k_{r}}=\frac{\rho\zeta P_{s}\left|\sum_{i=1}^{M}h_{i}g_{r_{i}}e^{-j\theta{}_{r}^{m}}\right|^{2}}{N_{0}d_{BV}^{\alpha}d_{VUr}^{\alpha}},
\end{equation}

\begin{equation}
\gamma_{k_{t}}=\frac{(1-\rho)\frac{(1-\zeta)P_{s}}{d_{BV}^{\alpha}d_{VU_{t}}^{\alpha}}\left|\sum_{i=1}^{M}h_{i}g_{t_{i}}e^{-j\theta{}_{t}^{m}}\right|^{2}}{\rho\frac{\zeta P_{s}}{d_{BV}^{\alpha}d_{VU_{t}}^{\alpha}}\left|\sum_{i=1}^{M}h_{i}g_{t_{i}}e^{-j\theta{}_{t}^{m}}\right|^{2}+N_{0}},
\end{equation}

\noindent and, the data rate of LU is $R_{k_{q}}=\log_{2}(1+\gamma_{k_{q}}).$

\noindent Moreover, the SINR at the eavesdroppers $\{e_{r},e_{t}\}$
for decoding the signal of BS using (4) and (5), can be given by:

\begin{equation}
\gamma_{e_{r}}=\frac{\rho\zeta P_{s}\left|\sum_{i=1}^{M}h_{i}g_{er_{i}}e^{-j\theta{}_{r}^{m}}\right|^{2}}{N_{0}d_{BV}^{\alpha}d_{VE_{r}}^{\alpha}},
\end{equation}

\begin{equation}
\gamma_{e_{t}}=\frac{(1-\rho)\frac{(1-\zeta)P_{s}}{d_{BV}^{\alpha}d_{VE_{t}}^{\alpha}}\left|\sum_{i=1}^{M}h_{i}g_{et_{i}}e^{-j\theta{}_{t}^{m}}\right|^{2}}{\rho\frac{\zeta P_{s}}{d_{BV}^{\alpha}d_{VE_{t}}^{\alpha}}\left|\sum_{i=1}^{M}h_{i}g_{et_{i}}e^{-j\theta{}_{t}^{m}}\right|^{2}+N_{0}}.
\end{equation}

\noindent Similarly, we can obtain the data rate of eavesdroppers
for decoding the signal of BS as $R_{e_{q}}=\log_{2}(1+\gamma_{e_{q}})$.
Accordingly, the secrecy rate is defined as the difference between the achievable data rate of each user and that of the eavesdroppers in both the reflection and transmission regions, which is given by:
\begin{equation}
R_{rn}^{sec}=\left[R_{k_{r}}-R_{_{rn_{e}}}\right]^{+},
\end{equation}
\begin{equation}
R_{tn}^{sec}=\left[R_{k_{t}}-R_{_{tn_{e}}}\right]^{+}.
\end{equation}

Therefore, the sum secure rate of a pair is 
\begin{equation}
R^{sec}=R_{rn}^{sec}+R_{tn}^{sec},
\end{equation}

where $[u]^{+}=\max\left\{ u,0\right\} .$

\subsection{SNR Distribution}

The signal-to-noise ratio (SNR) at the user LU is influenced by the
phase shifts of the reflecting elements of the RIS. The optimal phase
shift $\theta_{op}$ for the $n$-th reflecting element is given by:
\begin{equation}
\theta_{opt}=-(\theta_{h_{i}}+\theta_{g_{q_{i}}}),
\end{equation}

\noindent where $\theta_{h_{i}}$ and $\theta_{g_{q_{i}}}$are the
phases of the channels $h_{i}$ and $g_{q_{i}}$, respectively.

\subsubsection{Phase Estimation Errors}

Due to practical imperfections such as jittering and airflow effects,
the actual phase shift $\theta_{q}^{m}$ deviates from the optimal
phase:
\begin{equation}
\theta{}_{q}^{m}=\theta_{opt}+\phi.
\end{equation}

The phase estimation error $\phi$ follows a von-Mises distribution
with zero mean and concentration parameter $k$. The probability density
function (PDF) and eigen function of $\phi$ , respectively, are given
by
\begin{equation}
f_{\phi}(x)=\frac{e^{k\cos(x)}}{2\pi I_{0}(k)},
\end{equation}
\begin{equation}
\varphi_{p}=\mathbb{E}(e^{ip\phi})=\frac{I_{p}(k)}{I_{0}(k)}.
\end{equation}

\noindent where $I_{0}(.)$ and $I_{p}(.)$ are the modified Bessel
functions of the first kind of order zero and $p$, respectively.

\subsubsection{SNR Expressions}

Under phase estimation errors, the SNRs of users $\{k_{r}$, $k_{t}$\}
can be expressed as:

\begin{equation}
\gamma_{k_{r}}=\frac{\rho\zeta P_{s}\left|\sum_{i=1}^{M}h_{i}g_{r_{i}}e^{-j\theta_{r_{i}}}\right|^{2}}{N_{0}d_{BV}^{\alpha}d_{VUr}^{\alpha}}=k_{1}X_{r},
\end{equation}

\begin{align}
\gamma_{k_{t}} & =\frac{(1-\rho)\frac{(1-\zeta)P_{s}}{d_{BV}^{\alpha}d_{VU_{t}}^{\alpha}}\left|\sum_{i=1}^{M}h_{i}g_{t_{i}}e^{-j\theta_{t_{i}}}\right|^{2}}{\rho\frac{\zeta P_{s}}{d_{BV}^{\alpha}d_{VU_{r}}^{\alpha}}\left|\sum_{i=1}^{M}h_{i}g_{t_{i}}e^{-j\theta_{t_{i}}}\right|^{2}+N_{0}}\nonumber \\
 & =\frac{k_{2}X_{t}}{k_{1}X_{t}+1},
\end{align}

\noindent where $k_{1}=\frac{\rho\zeta P_{s}}{N_{0}d_{BV}^{\alpha}d_{VUr}^{\alpha}}$
and $k_{2}=\frac{(1-\rho)(1-\zeta)P_{s}}{N_{0}d_{BV}^{\alpha}d_{VUt}^{\alpha}}$,
with $P_{s}$ denoting the transmitting power BS, as previously stated, we employ Nakagami-m fading model to characterize the small-scale fading of A2G channels, due to its ability to represent a wide variety of multipath fading conditions by adjusting the fading severity parameter
$m$, where by denoting $X_{q}=\left|\sum_{i=1}^{M}\left|h_{i}\right|\left|g_{q_{i}}\right|e^{-j\phi_{i}}\right|^{2}=\left|U_{q}+jV_{q}\right|^{2}$
the expectations and variances of $U_{q}$ can be obtained as: 
\begin{align*}
\mathbb{E}(U_{r}) & =M\mathbb{E}(\left|h_{i}\right|\left|g_{r_{i}}\right|\cos\phi_{i})=M\alpha_{r}^{2}\varphi_{1},
\end{align*}
\begin{align*}
V(U_{r}) & =M\mathbb{E}(\left|h_{i}\right|\left|g_{r_{i}}\right|\cos\phi_{i})^{2}-M\mathbb{E}^{2}(\left|h_{i}\right|\left|g_{r_{i}}\right|\cos\phi_{i})\\
 & =\frac{M}{2}(1+\varphi_{2}-2\alpha_{r}^{4}\varphi_{1}^{2}),
\end{align*}
 where $\alpha_{q}=\sqrt{\mathbb{E}(\left|h_{i}\right|\mathbb{E}(\left|g_{q_{i}}\right|)}$,
$\mathbb{E}(\left|h_{i}\right|)=\frac{\Gamma(m_{BV}+0.5)}{\Gamma(m_{BV})\sqrt{m_{BV}}}$
and $\mathbb{E}(\left|g_{q_{i}}\right|)=\frac{\Gamma(m_{VU_{q}}+0.5)}{\Gamma(m_{VU_{q}})\sqrt{m_{VU_{q}}}}$.

$Proof:$ Please refer to Appendix A and B.

According to \cite{re47}, the PDF of $X_{q}$ is given by:
\begin{equation}
f_{X_{q}}(x)=\frac{x^{m_{BVU_{q}}-1}m_{BVU_{q}}{}^{m_{BVU_{q}}}}{\Gamma(m_{BVU_{q}})\Omega_{BVU_{q}}{}^{mBVU_{q}}}e^{-\frac{m_{BVU_{q}}}{\Omega_{BVU_{q}}}x}.
\end{equation}

\noindent where $m_{BVU}=\frac{\mathbb{E}^{2}(U_{q})}{4V(U_{q})}=\frac{M\alpha_{q}^{4}\varphi_{1}^{2}}{2(1+\varphi_{2}-2\alpha_{q}^{4}\varphi_{1}^{2})}$,
$\varOmega_{BVU_{r}}=\mathbb{E}^{2}(U_{q})=M^{2}\alpha_{q}^{4}\varphi_{1}^{2}$.

\noindent For the eavesdropper, the phase shifts are optimized with
respect to the legitimate user's channel, leading to a different effective
phase shift expression at the eavesdropper, which may cause signal
misalignment and reduce the received signal strength. It will no longer
depend on a single $\phi$, but it will include three key angular
components $\theta_{eff}=\theta_{g_{eq}}-\theta_{g_{q}}+\phi$ where,
$\phi$$\lyxmathsym{\AC}$von Mises distribution with zero mean and
concentration parameter $k$, $\{\theta_{g_{eq}},\theta_{g_{q}}\}$
are uniformly distributed over {[}0,2\textgreek{π}), representing
the unknown phases of the eavesdropper and legitimate user channels,
respectively. 

The composite error $\theta_{eff}$ can be approximated as a wrapped
normal random variable with zero mean and variance $Var(\theta_{eff})=Var(\theta_{g_{eq}})-Var(\theta_{g_{q}})+Var(\phi)$.
Considering Nakagami, $X_{e_{q}}=\left|\sum_{i=1}^{M}\left|h_{i}\right|\left|g_{e_{qi}}\right|e^{-j(\theta_{g_{er_{i}}}-\theta_{g_{r_{i}}}+\phi_{i})}\right|^{2}=\left|U_{e_{q}}+jV_{e_{q}}\right|^{2}.$

\noindent Therefore, the instantaneous SNRs for eavesdroppers in the
reflect and transmit areas are derived as follows:

\begin{equation}
\gamma_{rn_{e}}=\frac{\rho\zeta P_{s}\left|\sum_{i=1}^{M}h_{i}g_{er_{i}}e^{-j(\theta{}_{opt_{i}}+\theta_{eff_{i}})}\right|^{2}}{N_{0}d_{BV}^{\alpha}d_{VE_{r}}^{\alpha}}=k_{1}^{'}X_{er},
\end{equation}

\noindent where $k_{1}^{'}=\frac{\rho\zeta P_{s}}{N_{0}d_{BV}^{\alpha}d_{VE_{r}}^{\alpha}}$,
and $\begin{aligned}X_{er}=\left|\sum_{i=1}^{M}\left|h_{i}\right|\left|g_{er_{i}}\right|e^{-j\theta_{eff_{i}}}\right|^{2}\end{aligned}
$.

\begin{align}
\gamma_{tn_{e}} & =\frac{\frac{(1-\rho)(1-\zeta)P_{s}}{d_{BV}^{\alpha}d_{VE_{t}}^{\alpha}}\left|\sum_{i=1}^{M}h_{i}g_{et_{i}}e^{-j(\theta{}_{opt_{i}}+\theta_{eff_{i}})}\right|^{2}}{\frac{\rho\zeta P_{s}}{d_{BV}^{\alpha}d_{VE_{r}}^{\alpha}}\left|\sum_{i=1}^{M}h_{i}g_{et_{i}}e^{-j(\theta{}_{opt_{i}}+\theta_{eff_{i}})}\right|^{2}+N_{0}}\nonumber \\
 & =\frac{k_{2}^{'}X_{et}}{k_{1}^{'}X_{et}+1},
\end{align}

\noindent where $k_{1}^{'}=\frac{\rho\zeta P_{s}}{N_{0}d_{BV}^{\alpha}d_{VE_{r}}^{\alpha}}$,
$k_{2}^{'}=\frac{(1-\rho)(1-\zeta)P_{s}}{N_{0}d_{BV}^{\alpha}d_{VE_{t}}^{\alpha}}$.
The expectations and variances of $U_{e_{q}}$can be obtained as:
\[
\mathbb{E}(U_{er})=M\mathbb{E}(\left|h_{i}\right|\left|g_{er_{i}}\right|\cos(\theta_{eff_{i}}))=M\alpha_{eq}^{2}\Phi,
\]
\begin{align*}
V(U_{eq}) & =M\mathbb{E}(\left|h_{i}\right|\left|g_{eq_{i}}\right|\cos\theta_{eff_{i}})^{2}-M\mathbb{E}^{2}(\left|h_{i}\right|\left|g_{eq_{i}}\right|\cos\theta_{eff_{i}})\\
 & =\frac{M}{2}(1+\varphi_{2}-2\alpha_{eq}^{4}\Phi^{2}),
\end{align*}
 where $\Phi=e^{-\frac{Var(\theta_{eff_{i}})}{2}}$, $\alpha_{eq}=\sqrt{\mathbb{E}(\left|h_{i}\right|\mathbb{E}(\left|g_{eq_{i}}\right|)}$,
and $\mathbb{E}(\left|g_{eq_{i}}\right|)=\frac{\Gamma(m_{VE_{q}}+0.5)}{\Gamma(m_{VE_{q}})\sqrt{m_{VE_{q}}}}.$

Therefore, the PDF of $X_{er}$ is given by:
\begin{equation}
f_{X_{eq}}(x)=\frac{x^{m_{BVE_{q}}-1}m_{BVE_{q}}{}^{m_{BVE_{q}}}}{\Gamma(m_{BVE_{q}})\Omega_{BVE_{q}}{}^{mBVE_{q}}}e^{-\frac{m_{BVE_{q}}}{\Omega_{BVE_{q}}}x},
\end{equation}

where $m_{BVE_{q}}=\frac{M\alpha_{eq}^{4}\Phi^{2}}{2(1+\varphi_{2}-2\alpha_{eq}^{4}\Phi^{2})},$
and $\varOmega_{BVU_{r}}=\frac{M^{2}\alpha_{eq}^{4}\Phi^{2}}{m_{BVE_{q}}}$.

\section{PERFORMANCE ANALYSIS}

This section derives a precise expression for the ergodic secrecy rate applicable to both reflection and transmission regions in the STAR-RIS-assisted NOMA system, taking into account different UAV positions and power allocation schemes.
\subsection{Ergodic Secrecy Rate of the Transmit Area}

The ergodic capacity of the users $k_{t}$ at the transmit area is
expressed as:

\begin{spacing}{0}
\begin{center}
\begin{align}
C_{k_{t}} & =\frac{1}{\ln(2)}\mathbb{E}\left[\log_{2}(1+\gamma_{k_{t}})\right]\nonumber \\
 & =\frac{1}{\ln(2)}\mathbb{E}[\log_{2}(1+(\frac{k_{2}X_{t}}{k_{1}X_{t}+1})].\\
\nonumber 
\end{align}
\par\end{center}
\end{spacing}

Based on \cite{re48}, a moment generating function (MGF)-based expression
for the ergodic capacity is presented as:
\[
\mathbb{E}[\ln(\frac{k_{2}X_{t}}{k_{1}X_{t}+1})]
\]
\begin{align*}
= & \int_{0}^{\infty}\frac{\mathcal{M}_{\mathrm{X_{t}}}(k_{1}z)-\mathcal{M}_{\mathrm{X_{t}}}((k_{1}+k_{2})z)}{z}\mathrm{e^{-z}}\mathrm{d}z.
\end{align*}

It is noteworthy that $\mathcal{M_{\mathrm{X}}\mathrm{(\beta z)=\mathbb{E\mathrm{[\mathrm{\mathrm{e^{-z\beta X}]}}}}}}$
is the MGF of $\beta X$, $\beta\in\{k_{1},(k_{1}+k_{2}\}$. Therefore,
\begin{align}
\mathcal{M}_{\mathrm{X_{t}}}\mathrm{(k_{1}z)} & =E[\mathrm{\mathrm{e^{-zk_{1}X}]}=\int_{0}^{\infty}e^{-zk_{1}X}\mathit{f_{X_{t}}(x)}d\mathit{\mathit{x}}}\nonumber \\
 & =\int e^{-zk_{1}X}.\frac{x^{m_{BVU_{t}}-1}m_{BVU_{t}}{}^{m_{BVU_{t}}}}{\Gamma(m_{BVU_{t}})\Omega_{BVU_{t}}{}^{mBVU_{t}}}e^{-\frac{m_{BVU_{t}}}{\Omega_{BVU_{t}}}x}\mathrm{d}\mathit{\mathit{x}}\nonumber \\
 & =\left(\mathrm{1}+\frac{z\varOmega_{BVU_{t}}k_{1}}{m_{BVU_{t}}}\right)^{-m_{BVU_{t}}},
\end{align}

and
\[
\mathcal{M}_{\mathrm{X_{t}}}((k_{2}+k_{1})z)=\int_{0}^{\infty}e^{-z(k_{1}+k_{2})X}f_{X_{t}}(x)\mathrm{d\mathit{x}}
\]
\begin{align}
 & =\int_{0}^{\infty}e^{-z(k_{1}+k_{2})X}.\frac{x^{m_{BVU_{t}}-1}m_{BVU_{t}}{}^{m_{BVU_{t}}}}{\Gamma(m_{BVU_{t}})\Omega_{BVU_{t}}{}^{mBVU_{t}}}e^{-\frac{m_{BVU_{t}}}{\Omega_{BVU_{t}}}x}\mathrm{d\mathit{x}}\nonumber \\
 & =\left(1+\frac{z\varOmega_{BVU_{t}}(k_{2}+k_{1})}{m_{BVU_{t}}}\right)^{^{-m_{BVU_{t}}}}.
\end{align}

Therefore,

\[
C_{k_{t}}=\frac{1}{\ln(2)}\int_{0}^{\infty}\frac{e^{-z}}{z}\left[\mathbb{\mathrm{\textrm{(1+\ensuremath{\frac{z\varOmega_{BVU_{t}}k_{1}}{m_{BVU_{t}}})^{-m_{BVU_{t}}}}}}}\right.
\]

\begin{equation}
-\left.\left(1+\frac{z\varOmega_{BVU_{t}}(k_{2}+k_{1})}{m_{BVU_{t}}}\right)^{^{-m_{BVU_{t}}}}\right]\mathrm{d}z.
\end{equation}

Furthermore, the expression given in (27) can be written in terms
of the weights and sample points of the Laguerre orthogonal polynomial
as
\[
C_{k_{t}}=\frac{1}{\ln(2)}\sum_{i=1}^{N}\frac{w_{i}}{z_{i}}\left[\left(1+\frac{\varOmega_{BVU_{t}}\rho\zeta P_{s}z_{i}}{N_{0}m_{BVU_{t}}d_{BV}^{\alpha}d_{VU_{t}}^{\alpha}}\right)^{^{-m_{BVU_{t}}}}\right.
\]

\begin{equation}
\left.-\left[1+\frac{\varOmega_{BVU_{t}}z_{i}(1-\rho-\zeta+2\rho\zeta)P_{s}}{m_{BVU_{t}}N_{0}d_{BV}^{\alpha}d_{VU_{t}}^{\alpha}}\right]^{^{-m_{BVU_{t}}}}\right].
\end{equation}

Similarly, we can obtain the ergodic capacity of the eavesdropper
at the transmit area as follows
\[
C_{e_{t}}=\frac{1}{\ln(2)}\int_{0}^{\infty}\frac{\mathcal{M}_{\mathrm{X_{t}}}(k_{1}^{'}z)-\mathcal{M}_{\mathrm{X_{t}}}((k_{1}^{'}+k_{2}^{'})z)}{z}\mathrm{e^{-z}}\mathrm{d}z,
\]
where 
\begin{align}
\mathcal{M}_{\mathrm{X_{e}}}\mathrm{(k_{1}^{'}z)}= & \mathbb{E}\mathrm{[\mathrm{\mathrm{e^{-zk_{1}X_{e}}]}}}\nonumber \\
= & \int_{0}^{\infty}e^{-zk_{1}^{'}X_{e}}\mathit{f_{X_{et}}(x)}\mathrm{d}x\nonumber \\
= & \int e^{-zk_{1}^{'}X_{e}}\frac{x^{m_{BVE_{t}}-1}m_{BVE_{t}}{}^{m_{BVE_{t}}}}{\Gamma(m_{BVE_{t}})\Omega_{BVE_{t}}{}^{mBVE_{t}}}e^{-\frac{m_{BVE_{t}}}{\Omega_{BVE_{t}}}x}\mathrm{d}x\nonumber \\
= & \textrm{\ensuremath{\left(1+\frac{z\varOmega_{BVE_{t}}k_{1}^{'}}{m_{BVE_{t}}}\right)^{-m_{BVE_{t}}},}}
\end{align}

and
\begin{align*}
\mathcal{M}_{\mathrm{X_{t}}}((k_{2}^{'}+k_{1}^{'})z)= & \int_{0}^{\infty}e^{-z(k_{1}^{'}+k_{2}^{'})X_{e}}f_{X_{et}}(x)\mathrm{d}x
\end{align*}
\begin{align}
= & \int_{0}^{\infty}e^{-z(k_{1}+k_{2})X_{e}}\frac{x^{m_{BVE_{t}}-1}m_{BVE_{t}}{}^{m_{BVE_{t}}}}{\Gamma(m_{BVE_{t}})\Omega_{BVE_{t}}{}^{mBVE_{t}}}e^{-\frac{m_{BVE_{t}}}{\Omega_{BVE_{t}}}x}\mathrm{d}x\nonumber \\
= & \left(1+\frac{z\varOmega_{BVE_{t}}(k_{1}^{'}+k_{2}^{'}}{m_{BVE_{t}}}\right)^{-m_{BVE_{t}}}.
\end{align}

Therefore, 
\[
C_{e_{t}}=\frac{1}{\ln(2)}\int_{0}^{\infty}\frac{\mathbb{\boldsymbol{\mathrm{1}}}}{z}e^{-z}\left[\mathfrak{\mathcal{\mathrm{\mathit{\left(\mathrm{1}+\frac{z\varOmega_{BVE_{t}}k_{1}^{'}}{m_{BVE_{t}}}\right)^{^{-m_{BVE_{t}}}}}}}}\right.
\]
\begin{equation}
\left.-\mathcal{\mathrm{\mathit{\left(\mathrm{1}+\frac{z\varOmega_{BVE_{t}}\mathrm{(}k_{1}^{'}+k_{2}^{'}\mathrm{)}}{m_{BVE_{t}}}\right)^{^{-m_{BVE_{t}}}}}}}\right]\mathrm{d}z.
\end{equation}
The expression given in (27), can be written in terms of the weights
and sample points of the Laguerre orthogonal polynomial as

\begin{spacing}{0}
\begin{center}
\[
C_{e_{t}}=\frac{1}{\ln(2)}\sum_{i=1}^{N}\frac{w_{i}}{z_{i}}\left(\left(1+\frac{\varOmega_{BVE_{t}}\rho\zeta P_{s}z_{i}}{N_{0}m_{BVE_{t}}d_{BV}^{\alpha}d_{VE_{t}}^{\alpha}}\right)^{^{-m_{BVE_{t}}}}\right.
\]
\par\end{center}
\end{spacing}

\begin{equation}
\left.-\left(1+\frac{\varOmega_{BVE_{t}}z_{i}(1-\rho-\zeta+2\rho\zeta)P_{s}}{m_{BVE_{t}N_{0}d_{BV}^{\alpha}d_{VE_{t}}^{\alpha}}}\right)^{^{-m_{BVE_{t}}}}\right).
\end{equation}

Based on (12) the secrecy rate can be obtain as:
\[
R_{t}^{sec}=\left[\frac{1}{\ln(2)}\sum_{i=1}^{N}\frac{w_{i}}{z_{i}}\left(\left(1+\frac{\varOmega_{BVU_{t}}\rho\zeta P_{s}z_{i}}{N_{0}m_{BVU_{t}}d_{BV}^{\alpha}d_{VU_{t}}^{\alpha}}\right)^{^{-m_{BVU_{t}}}}\right.\right.
\]

\[
-\left(1+\frac{\varOmega_{BVU_{t}}z_{i}(1-\rho-\zeta+2\rho\zeta)P_{s}}{m_{BVU_{t}}N_{0}d_{BV}^{\alpha}d_{VU_{t}}^{\alpha}}\right)^{^{-m_{BVU_{t}}}}
\]

\[
-\left(1+\frac{\varOmega_{BVE_{t}}\rho\zeta P_{s}z_{i}}{N_{0}m_{BVE_{t}}d_{BV}^{\alpha}d_{VE_{t}}^{\alpha}}\right)^{^{-m_{BVE_{t}}}}
\]

\begin{equation}
\left.+\left(1+\frac{\varOmega_{BVE_{t}}z_{i}(1-\rho-\zeta+2\rho\zeta)P_{s}}{m_{BVE_{t}}N_{0}d_{BV}^{\alpha}d_{VE_{t}}^{\alpha}}\right)\right]^{+}.
\end{equation}

\subsection{Ergodic Secrecy Rate of the Reflect Area}

The ergodic capacity of the user $k_{r}$ at the reflect area is expressed
as:

\begin{align}
C_{k_{r}} & =\frac{1}{\ln(2)}\mathbb{E}\left[\log_{2}(1+\gamma_{k_{r}})\right]=\frac{1}{\ln(2)}\mathbb{E}[\ln(1+k_{1}X_{r})]\nonumber \\
 & =\frac{1}{\ln(2)}\mathbb{E}\left[\log_{2}\left(1+\frac{\rho\zeta P_{s}\left|\sum_{i=1}^{M}h_{i}g_{r_{i}}e^{-j\phi_{i}}\right|^{2}}{N_{0}d_{BV}^{\alpha}d_{VUr}^{\alpha}}\right)\right]\nonumber \\
 & =\frac{1}{\ln(2)}\int_{0}^{\infty}\frac{\mathbb{\textrm{1-}\mathfrak{\mathcal{M_{\mathrm{X_{r}}}\mathit{\mathrm{(}z\mathrm{)}}}}}}{z}e^{-z}\mathrm{d}z\nonumber \\
 & =\frac{1}{\ln(2)}\int_{0}^{\infty}\frac{1-\left(1+\frac{z\varOmega_{BVU_{r}}k_{1}}{m_{BVU_{r}}}\right)^{^{-m_{BVU_{r}}}}}{z}e^{-z}\mathrm{d}z,
\end{align}

\noindent where $M_{\mathrm{X_{r}}}(k_{1}z)=(1+\frac{z\varOmega_{BVU_{r}}k_{1}}{m_{BVU_{r}}})^{-m_{BVU_{r}}}$,
the expression given in (30), can be written as
\begin{align}
C_{k_{r}} & =\frac{1}{\ln(2)}\sum_{i=1}^{N}\frac{w_{i}}{z_{i}}\nonumber \\
 & \,\,\,\,\left[1\left.-\left(1+\frac{\varOmega_{BVU_{r}}\rho\zeta P_{s}z_{i}}{N_{0}m_{BVU_{r}}d_{BV}^{\alpha}d_{VU_{r}}^{\alpha}}\right)^{^{-m_{BVU_{r}}}}\right].\right.
\end{align}

Similarly,we can obtain the ergodic capacity of the eavesdropper as
follows
\begin{align}
C_{e_{r}} & =\frac{1}{\ln(2)}\mathbb{E}\left[\log_{2}(1+\gamma_{e_{r}})\right]=\frac{1}{\ln(2)}\mathbb{E}[\ln(1+k_{1}^{'}X_{er})]\nonumber \\
 & =\frac{1}{\ln(2)}\int_{0}^{\infty}\frac{\mathbb{\textrm{1-}\mathfrak{\mathcal{M_{\mathrm{X_{er}}}\textrm{\ensuremath{\mathit{\mathrm{(}z\mathrm{)}}}}}}}}{z}e^{-z}\mathrm{d}\mathit{z}\nonumber \\
 & =\frac{1}{\ln(2)}\int_{0}^{\infty}\frac{1-\left(1+\frac{z\varOmega_{BVE_{r}}k_{1}^{'}}{m_{BVE_{r}}}\right)^{^{-m_{BVE_{r}}}}}{z}e^{-z}\mathrm{d}z.
\end{align}

Therefore, the expression given in (36) can be written as:
\begin{align}
C_{e_{r}} & =\frac{1}{\ln(2)}\sum_{i=1}^{N}\frac{w_{i}}{z_{i}}\nonumber \\
 & \,\,\,\,\left[1\left.-\left(1+\frac{\varOmega_{BVE_{r}}\rho\zeta P_{s}z_{i}}{N_{0}m_{BVE_{r}}d_{BV}^{\alpha}d_{VE_{r}}^{\alpha}}\right)^{^{-m_{BVE_{r}}}}\right].\right.
\end{align}

Based on (11) the secrecy rate can be obtained as:

\begin{spacing}{0}
\begin{center}
\[
R_{r}^{sec}=\left[\frac{1}{ln(2)}\sum_{i=1}^{N}\frac{w_{i}}{z_{i}}\left(\left(1+\frac{\varOmega_{BVE_{r}}\rho\zeta P_{s}z_{i}}{N_{0}m_{BVE_{r}}d_{BV}^{\alpha}d_{VE_{r}}^{\alpha}}\right)^{^{-m_{BVE_{r}}}}\right.\right.
\]
\begin{equation}
\left.\left.-\left(1+\frac{\varOmega_{BVU_{r}}\rho\zeta P_{s}z_{i}}{N_{0}m_{BVU_{r}}d_{BV}^{\alpha}d_{VU_{r}}^{\alpha}}\right)^{^{-m_{BVU_{r}}}}\right)\right]^{+}.
\end{equation}
\par\end{center}
\end{spacing}

\section{JOINT OPTIMIZATION OF WSSR}

In this section, we adopt an alternating optimization (AO) approach
to solve the non-convex joint optimization problem in (39), which
involves two key variables: the UAV’s 3D placement and the power allocation
ratio of the STAR-RIS. Specifically, the original problem is decomposed
into two subproblems, each solved iteratively while fixing the other
variable. For the UAV placement subproblem, which is non-convex due
to the complex dependence of the secrecy rate on the UAV's position,
we employ the 3D search alogrithim method to obtain an efficient solution.
For the second subproblem involving the non-convex STAR-RIS power
allocation ratio, we apply a one-dimensional search or approximation-based
method to determine the optimal power split between the transmission
and reflection regions. 

\begin{subequations}
\begin{align}
\underset{C_{u},\zeta}{\mathbf{\max\,\,\,}} & w_{1}R_{t}^{sec}+w_{2}R_{r}^{sec}\\
s.t.\,\,\, & w_{1}+w_{2}=1,w_{1},w_{2}\in[0,1],w_{1}<w_{2}\\
 & x_{min}\le x_{u}\le x_{max}\\
 & y_{min}\le y_{u}\le y_{max}\\
 & z_{min}\le z_{u}\le z_{max}\\
 & \zeta_{t}^{m}+\zeta_{r}^{m}=1,\zeta_{q}^{m}\in[0,1],q\in[r,t],\forall m.
\end{align}
\end{subequations}

\subsection{Optimization of UAV Placement}

In this subsection, we optimize the UAV’s 3D positioning through linear
grid-based search algorithm where two coordinates are held fixed while
the third is adjusted iteratively. This strategy ensures systematic
convergence toward an optimal location that maximizes the secrecy
rate. Therefore, given the power allocation $\zeta$ the optimization
problem of UAV position can be separated from (39) as:

\begin{subequations}
\begin{align}
\underset{C_{u}}{\max\,\,\,} & w_{1}R_{t}^{sec}+w_{2}R_{r}^{sec}\\
s.t.\,\,\, & w_{1}+w_{2}=1,w_{1},w_{2}\in[0,1],w_{1}<w_{2}\\
 & x_{min}\le x_{u}\le x_{max}\\
 & y_{min}\le y_{u}\le y_{max}\\
 & z_{min}\le z_{u}\le z_{max}.
\end{align}
 
\end{subequations}

\noindent In this subsection, we focus on optimizing the 3D placement
of the UAV to maximize the WSSR under a given power allocation configuration
$\zeta$ of the STAR-RIS. However, the problem is inherently non-convex,
as the secrecy rate expressions are complex and non-linear functions
of the UAV's coordinates.

\noindent To efficiently tackle this challenge, we adopt a grid-based
iterative coordinate descent approach, where at each step, two coordinates
are held fixed while the third is optimized over a discretized search
space. This process is cyclically repeated over the three dimensions
until is reached. Such a strategy enables systematic exploration of
the 3D position space and provides a practical solution with reasonable
computational complexity. The iterative algorithm to solve this subproblem
is detailed in Algorithm 1.
\begin{algorithm}
\caption{Iterative algorthim for problem (40)}

1. Initial UAV position $x_{u},y_{u},z_{u}$, where $x=x_{\min}:1:x_{\max}$,
$y=y_{\min}:1:y_{\max}$, $z=z_{\min}:1:z_{\max}$. Set iteration
index $k=1$, and convergence threshold $\epsilon_{0}>0$.\\
2. Fix $x_{u}^{(k)}$, $y_{u}^{(k)}$, search over $z_{u}\in[z_{\min},z_{\max}]$.\\
3. Evaluate the objective $w_{1}R_{t}^{sec}+w_{2}R_{r}^{sec}$.\\
4. Select $z_{u}^{(k+1)}$ that maximizes the WSSR.\\
5. Fix $z_{u}^{(k+1)}$ ,$y_{u}^{(k)}$, search over $x_{u}\in[x_{\min},x_{\max}]$.\\
6. Repeat the evaluation and update $x_{u}^{(k+1)}$.\\
7. Fix $z_{u}^{(k+1)}$, $x_{u}^{(k+1)}$, search over $y_{u}\in[y_{\min},y_{\max}]$.\\
8. Repeat the evaluation and update $y_{u}^{(k+1)}$.\\
9. Check convergence:$\left\Vert \mathbf{C}_{u}^{(k+1)}-\mathbf{C}_{u}^{(k)}\right\Vert <\epsilon_{0}$.\\
10. Break.\\
11. Set $k\gets k+1$ Until $\{\ensuremath{k=K_{\max}\}}$.\\
12. Return.\\
13. $\ensuremath{(x_{u}^{*},y_{u}^{*},z_{u}^{*})=\mathbf{C}_{u}^{(k+1)}}$.
\end{algorithm}

\subsection{Optimization of Power Allocation}

Given the optimized UAV position $C_{u}$, we aim to further enhance
the secrecy performance of the system while ensuring fairness between
users. To this end, we adopt a power allocation strategy that maximizes
the WSSR. The corresponding optimization problem is defined as follows:

\begin{subequations}
\begin{align}
\underset{\zeta}{\max\,\,\,} & w_{1}R_{t}^{sec}+w_{2}R_{r}^{sec}\\
s.t.\,\,\, & w_{1}+w_{2}=1,w_{1},w_{2}\in[0,1],w_{1}<w_{2}\\
 & \zeta_{t}^{m}+\zeta_{r}^{m}=1,\zeta_{q}^{m}\in[0,1],q\in[r,t],\forall m.
\end{align}
\end{subequations}

\noindent To address the non-convexity of the WSSR function with respect
to $\zeta$, we employ the Golden Section Search (GSS) algorithm.
This method is highly suitable for solving one-dimensional unimodal
optimization problems, offering a balance between accuracy and computational
efficiency. The GSS technique iteratively narrows the search interval
for $\zeta$, converging to the optimal value that maximizes the WSSR
under the given fairness and feasibility constraints. The iterative
algorithm to solve this subproblem is detailed in Algorithm 2.

\begin{algorithm}
\caption{GSS for Optimal Power Allocation}

\begin{raggedright}
1. Initial search interval $[a,b]=[0,1]$, tolerance $\varepsilon>0$,
weights $w_{1},w_{2}$ satisfying $w_{1}+w_{2}=1$ and $w_{1}<w_{2}$,
golden ratio $\varphi=(\surd5-1)/2\approx0.618$, Set iteration index
$n$ = 0.\\
2. Compute initial points $c=b-\varphi(b-a)$,$d=a+\varphi(b-a).$\\
3. Evaluate WSSR at $c$ and $d$ : $f_{c}=w_{1}R_{t}^{sec}(c)+w_{2}R_{r}^{sec}(c)$,
$f_{d}=w_{1}R_{t}^{sec}(d)+w_{2}R_{r}^{sec}(d).$\\
4. While ($b-a>\varepsilon$) and ($n<N_{\max}$):\\
5. If $f_{c}>f_{d}$:
\par\end{raggedright}
6. Set $b=d$, $d=c$, and $c=b-\varphi(b-a)$,

7. Update $f_{d}=f_{c}$.\\
8. Compute $f_{c}=w_{1}R_{t}^{sec}(c)+w_{2}R_{r}^{sec}(c)$\\
9. Else:

10. Set $a=c$, $c=d$ and $d=a+\varphi(b-a)$.\\
11. Update $f_{c}=f_{d}$.

12. Compute $f_{d}=w_{1}R_{t}^{sec}(d)+w_{2}R_{r}^{sec}(d)$.\\
13. Set $n\gets n+1.$\\
14. Return optimal $\zeta^{*}=(a+b)/2$.
\end{algorithm}

\section{NUMERICAL RESULTS}

In this section, we investigate a wireless communication system comprising
three legitimate users and three eavesdroppers operating in both reflect
and transmit areas. The BS is positioned at coordinates (5, 5, 5),
while legitimate users in the reflect area are located at (1, 1, 0),
(0.5, 0.5, 0), (2, 0.5, 0), and those in the transmit area at (-1,
-1, 0), (-0.5, -0.5, 0), (-2, -0.5, 0). Eavesdroppers are placed at
(2, 2, 0), (0.25, 0.25, 0), (2, 1.5, 0), in the reflect area and (-2,
-2, 0), (-0.25, -0.25, 0), (-2, -1.5, 0) in the transmit area, and
the UAV is positioned at coordinates (0.5, 0.5, 10). The system employs
Nakagami-$m$ fading channels with a severity parameter of $m=2$
for all links. In the reflect region, the STAR-RIS power coefficient
$\zeta$ is set to 0.2, indicating 20\% power allocated for reflection,
while the NOMA power-splitting coefficient $\rho$ is set to 0.3,
allocating 30\% of the transmit power to users in the reflect area.
To verify our analysis, Monte Carlo simulation results are provided.
Secrecy performance is evaluated across transmit power levels ranging
from 0 to 50 dBm. This analysis offers key insights into the power-security
trade-offs for dual-area systems with spatially distributed adversaries,
highlighting the critical role of directional interference management
in enhancing PLS.

Fig. 2 shows the secrecy rate versus transmit power in the transmit
area of a STAR-RIS-assisted UAV NOMA system under different values
of the concentration parameter $k$ of the von Mises distribution.
The plot includes both analytical and Monte Carlo (MC) simulation
results for $k=10,15,20$, where a higher $k$ value indicates lower
phase uncertainty and more accurate phase alignment. As the transmit
power increases, the secrecy rate initially improves due to enhanced
signal quality at the legitimate user compared to the eavesdropper.
However, beyond a certain power level (around 25--50 dBm), the secrecy
rate begins to decline. This is because high transmit power also benefits
the eavesdropper and amplifies the negative effects of residual phase
errors and inter-user interference in the NOMA system. Moreover, lower
$k$ values (e.g., $k=10$) correspond to higher phase uncertainty,
which reduces beamforming accuracy and degrades secrecy performance.
In contrast, higher $k$ values (e.g., $k=20$) lead to more focused
phase distributions, resulting in improved secrecy rates. 
\begin{figure}
\raggedright
\includegraphics[bb=0bp 0bp 420bp 315bp,scale=0.4]{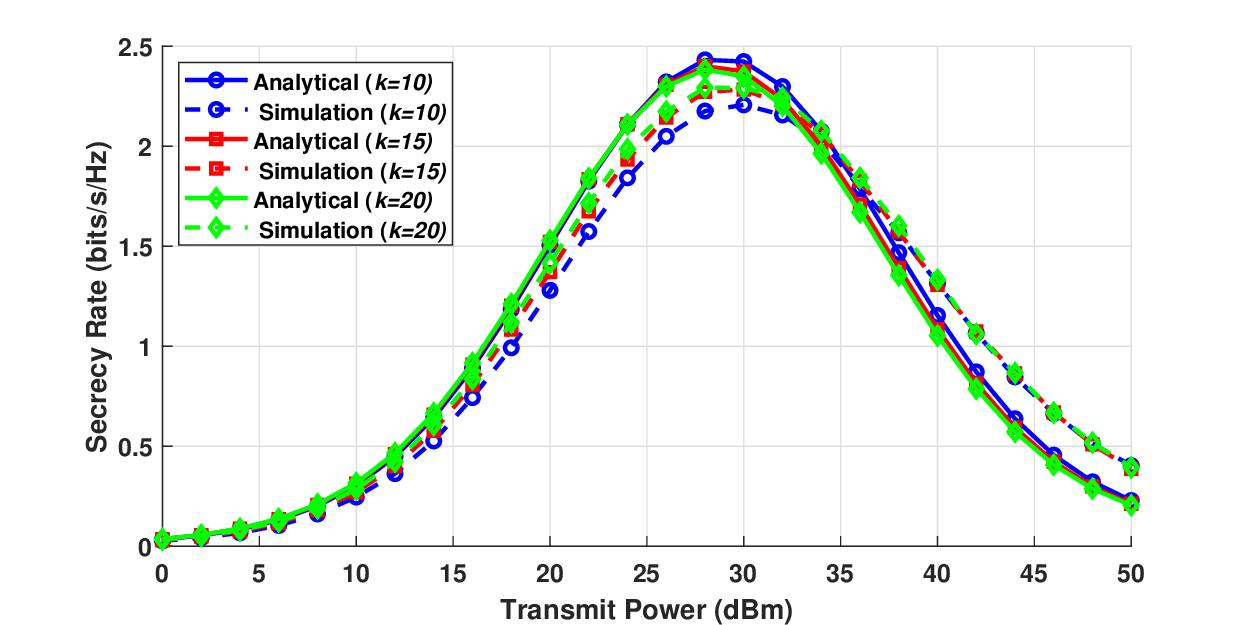}\caption{The secrecy rate versus the transmit power for the transmit area with
different values of $\kappa.$}
\end{figure}

Unlike in the transmit area, the secrecy rate in Fig. 3 monotonically
increases with transmit power and does not peak or decline. This trend
indicates that, in the reflect area, the benefit of increased transmit
power consistently outweighs the negative effects of inter-user interference
and phase estimation errors. As the transmit power increases, the SNR
at the legitimate user improves significantly, leading to a higher
achievable rate. The eavesdropper, however, experiences only limited
gain from increased power due to the suboptimal STAR-RIS alignment
and less favorable propagation conditions in the reflect region. This
asymmetric improvement leads to a steadily growing secrecy rate. Moreover,
as the von Mises concentration parameter $k$ increases implying lower
phase noise and more accurate RIS phase control the system achieves
better beamforming precision. Consequently, the secrecy rate improves
with higher $k$. The reflect area exhibits a more gradual improvement
in its secrecy rate $R_{r}^{sec}$. Although there is an upward trend,
the rate of increase is noticeably slower compared to the transmit
area. This difference can be attributed to the inherent characteristics
of the reflect area, where additional challenges such as multipath
propagation and reduced direct signal strength can impede performance.
The eavesdropper's capacity in this environment improves at a different
rate than that of the legitimate user, leading to a less pronounced
enhancement in secrecy rates. Adjusting the value $\rho=0.3$ directly
influences the power allocation between the reflect and transmit areas.
Specifically, $\rho$ determines the fraction of power allocated to
the reflect area, while ($1-\rho$) is allocated to the transmit area.
A lower value of $\rho$ (that is, allocating less power to the reflect
area and more to the transmit area) enhances the performance in the
transmit region, thereby increasing the secrecy rate due to better
signal quality for legitimate users and reduced exposure to eavesdroppers.
While the transmit area consistently outperforms the reflect area
in the low-to-medium power range due to favorable power allocation
and propagation conditions, the performance gap narrows as the power
increases. This suggests that under high power conditions, the reflect
area will begin to show improvements in secrecy performance that reduce
the dominance of the transmit area. Therefore, increasing $\rho$
allocates more power to the reflect area, that will increase $R_{r}^{sec}$
and reduce $R_{t}^{sec}$, especially at higher power levels.

\begin{figure}
\raggedright
\includegraphics[bb=0bp 0bp 420bp 315bp,scale=0.4]{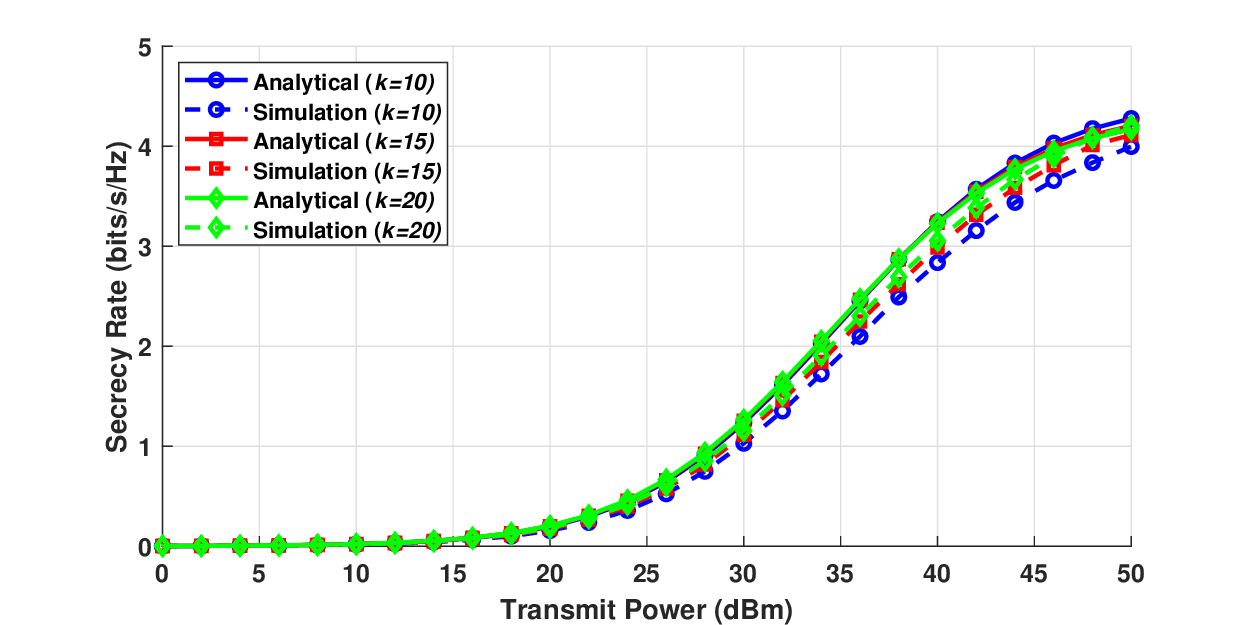}

\caption{The secrecy rate versus the transmit power for the reflect area with
different values of $\kappa$.}
\end{figure}

In Fig. 4, $R_{t}^{sec}$ is analyzed as a function of the number
of RIS elements $M$ at specific transmit power levels of 10 dBm,
15 dBm. The results indicate a clear relationship where increasing
the number of RIS elements leads to substantial enhancements in achievable
secrecy rates. For instance, as $M$ increases from 10 to 100, the
secrecy rate significantly improves, particularly at higher power
levels. It also apparent that the Monte Carlo simulations validate
the analytical curves, underscoring the effectiveness of deploying
additional RIS elements to boost the system's secrecy capacity.

\begin{figure}
\raggedright
\includegraphics[bb=0bp 0bp 420bp 315bp,scale=0.4]{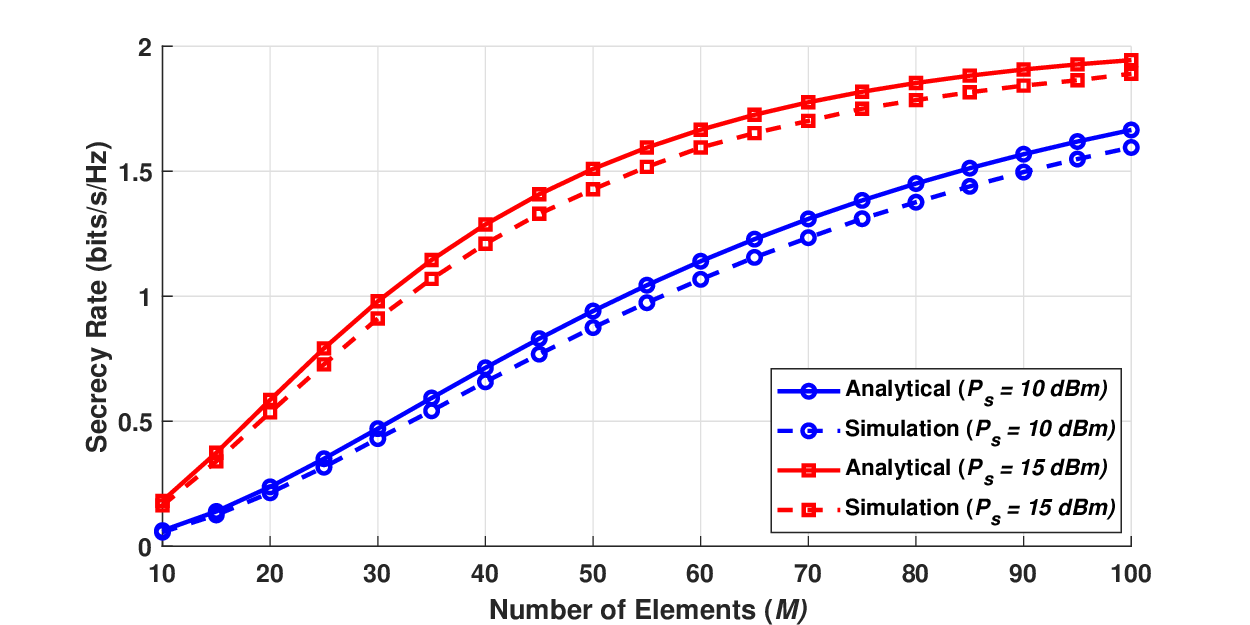}

\caption{The secrecy rate versus the number of the elements $M$ for the transmit
area with different transmit power values.}
\end{figure}

Similarly, Fig. 5 presents the secrecy rate performance in the reflect
area $R_{r}^{sec}$ concerning different values of $M$ with the same
transmit power levels. The trends observed indicate a more gradual
increase in secrecy rates compared to the transmit area. While the
increase is significant, it is not as pronounced as in the transmit
area. This analysis confirms that increasing the number of RIS elements
can significantly enhance PLS, even in the reflect area, albeit at
a slower rate.

\begin{figure}
\raggedright
\includegraphics[bb=0bp 0bp 420bp 315bp,scale=0.4]{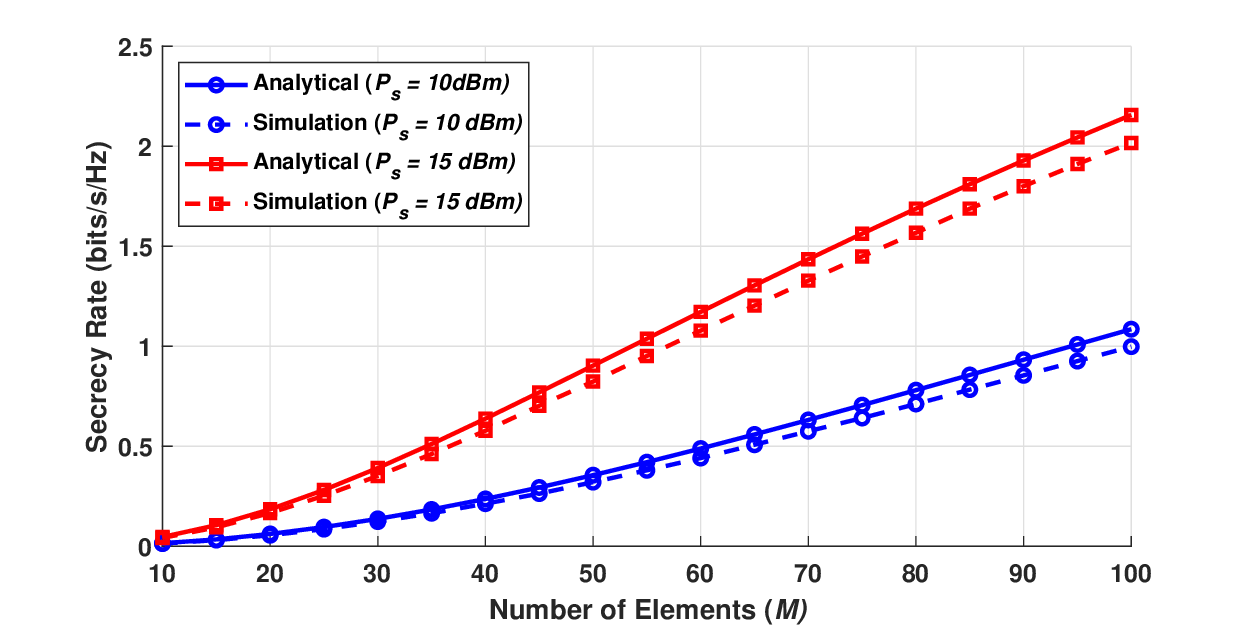}

\caption{The secrecy rate versus the number of the elements $M$ for the transmit
area with different transmit power values.}
\end{figure}

To maximize the total weighted secrecy rate in the proposed UAV-mounted
STAR-RIS-assisted NOMA system, we first investigate the impact of
the power-splitting ratio $\zeta.$ The optimization process is carried
out using Algorithm 2, which iteratively searches for the value of
$\zeta$ that yields the highest total weighted secrecy rate under
the given system configuration.

As depicted in Fig. 6, the WSSR initially increases with $\zeta$,
reaching its maximum at $\zeta^{*}=0.15$. Beyond this point, further
increases in $\zeta$ lead to a slight degradation in performance,
attributed to the imbalance between reflection and transmission energy
allocation in the STAR-RIS structure. Thus, $\zeta^{*}=0.15$ is selected
as the optimal value for the subsequent evaluations.

\begin{figure}[t]
\centering
\includegraphics[scale=0.5]{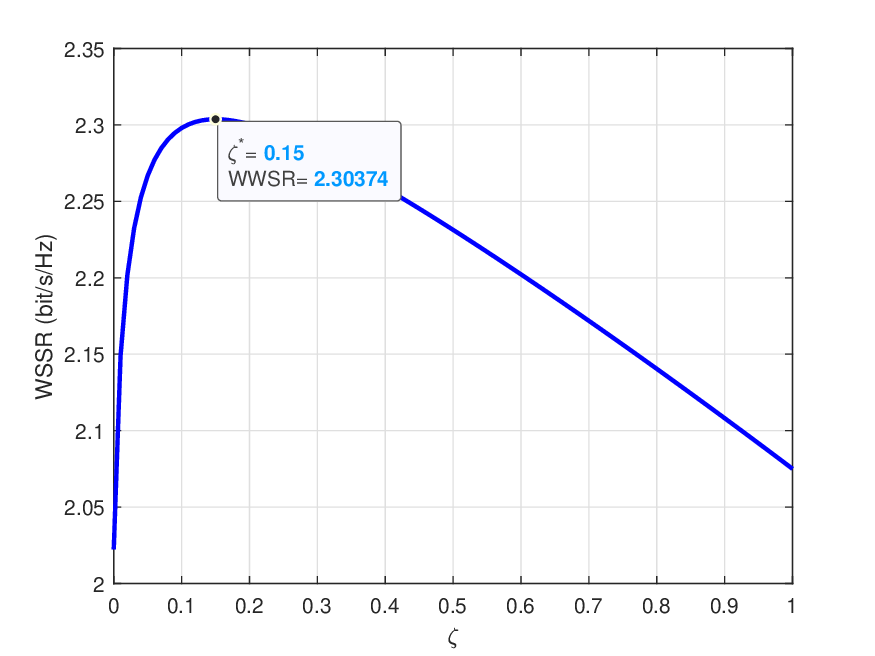}

\caption{WSSR as a function of energy allocation ratio $\zeta$ in the STAR-RIS-assisted
system.}
\end{figure}

Therefore, to evaluate the system’s secrecy performance under different
fairness strategies, the WSSR is evaluated as a function of the transmit
power under two distinct weighting configurations. As shown in Fig.
7, the WSSR is evaluated under a biased fairness scenario where $w_{1}=0.45$
and $w_{2}=0.55$, favoring the strong user. As observed, this configuration
yields a higher WSSR across the transmit power range, particularly
in the high-power regime, due to the stronger user's channel conditions
being given more weight in the optimization process. 

In contrast, Fig. 8 illustrates the performance under an equal fairness
setting, with $w_{1}=w_{2}=0.5$. Although the overall WSSR is slightly
lower compared to the biased configuration, this setup ensures a more
balanced secrecy performance between the two users, which is desirable
in fairness-constrained scenarios. For both scenarios, the system
parameters are $M=20$, $k=20$ and $\rho=0.3$.

\begin{figure}[t]
\centering
\includegraphics[scale=0.5]{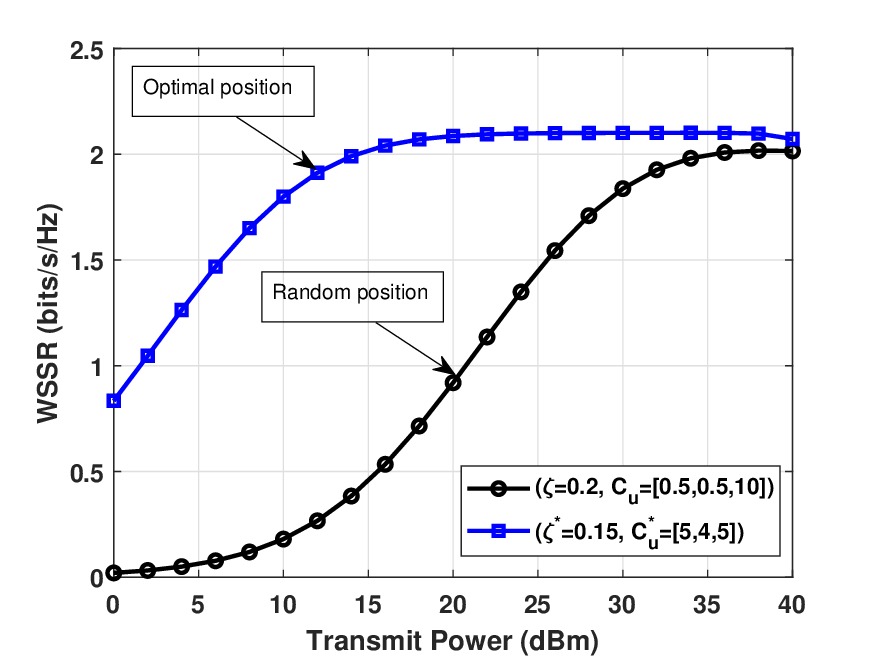}

\caption{WSSR versus transmit power under a biased fairness scenario $w_{1}<w_{2}.$}
\end{figure}

\begin{figure}[t]
\centering
\includegraphics[scale=0.5]{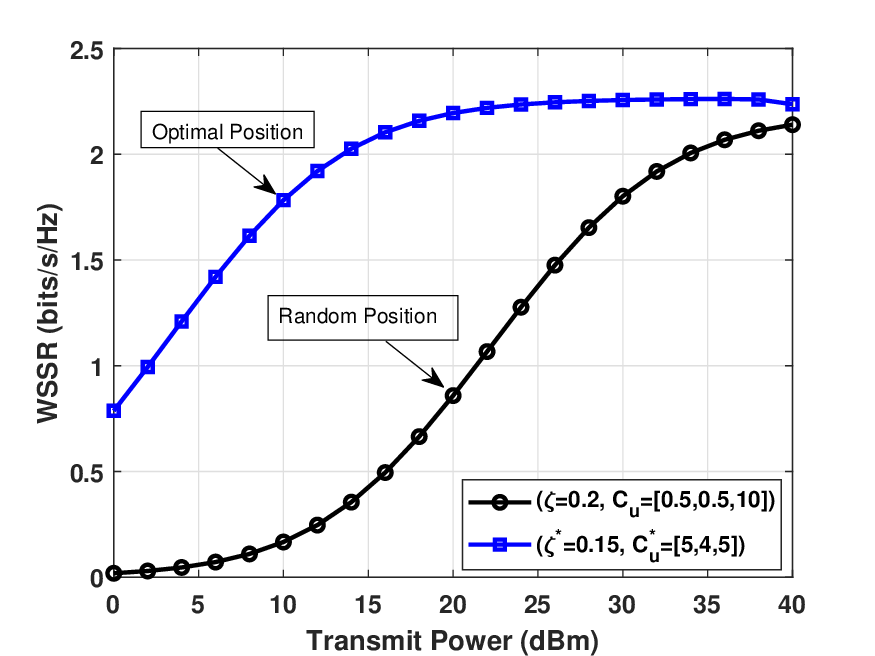}

\caption{Total secrecy rate versus transmit power under an equal fairness scenario
$w_{1}=w_{2}.$}

\end{figure}

\section{CONCLUSION}

In this work, we presented a UAV-mounted STAR-RIS-assisted NOMA system
for secure downlink communication in the presence of passive eavesdroppers.
By modeling the phase estimation error using a von Mises distribution
and approximating the composite angular error with a wrapped normal
distribution, we derived accurate closed-form expressions for the
ergodic secrecy rates in both reflection and transmission regions.
Results show that the proposed analytical expressions closely match
the Monte Carlo simulations, validating their accuracy and computational
efficiency. In addition, the results demonstrate that the joint optimization
of UAV placement and STAR-RIS power splitting significantly enhances
secrecy performance, especially under high transmit power. Moreover,
the findings reveal that phase estimation errors due to UAV mobility
introduce noticeable degradation, and it was also shown that increasing
the number of STAR-RIS elements improves secrecy but with diminishing
returns. Finally, the optimal configuration is achieved when the power-splitting
ratio and UAV location are jointly tuned, yielding the maximum total
secrecy rate in our considered system.

These insights not only validate the robustness of the proposed model
but also pave the way for the practical deployment of STAR-RIS-assisted
UAV NOMA systems in next-generation secure wireless networks.

\section*{APPENDIX A}

The $M$ complex variables $\left|g_{r}\right|\left|h\right|$e$^{j\phi}$,
$m$=1,..., $M$, are i.i.d. with common mean $\mu=\alpha_{r}^{2}\varphi_{1},$
variance $\upsilon=1-\alpha^{4}\left|\varphi_{1}^{2}\right|$, and
pseudo-variance $\rho=\varphi_{2}-\alpha^{4}\varphi_{1}^{2}$ (see\cite{re51}
for the second-order characterization of complex variables). 

According to the assumptions on the distribution of $\phi$, the trigonometric
moments $\varphi_{1},\varphi_{2}\epsilon\mathbb{R}$ and, consequently
$\mu,\rho\epsilon\mathbb{R}$. By invoking the central limit theorem,
we approximate the distribution of $\sqrt{X}$ for $M$ large by a
complex normal distribution $\mathcal{CN}(\mu,\upsilon/n,\rho/n)$.
Let $\mathit{U}=\mathfrak{R}(H)$and $V$= $\mathfrak{\Im}(H).$ Based
on {[}45\nocite{re51}{]}, we have
\begin{equation}
\mathrm{Cov}\left[\mathit{U,V}\right]=\frac{1}{2}\mathfrak{\Im}(-\frac{v}{n}+\frac{\rho}{n})=0.
\end{equation}

Being jointly Gaussian with zero covariance, it follows that $\mathit{U}$
and $\mathit{V}$ are independent. Furthermore, $\mathit{U}$$\sim$$\mathcal{N}(\mu,\sigma_{U}^{2})$,
$\mathit{V}\sim\mathcal{N}(0,\sigma_{V}^{2})$ and $\mu=\alpha^{2}\varphi_{1}$

\begin{equation}
\sigma_{U}^{2}=\frac{1}{2}\mathbb{\mathfrak{R}}(\frac{v}{n}+\frac{\rho}{n})=\frac{1}{2n}(1+\varphi_{2}-2\alpha^{4}\varphi_{1}^{2}),
\end{equation}
\begin{equation}
\sigma_{V}^{2}=\frac{1}{2}\mathbb{\mathfrak{R}}(\frac{v}{n}-\frac{\rho}{n})=\frac{1}{2n}(1-\varphi_{2}).
\end{equation}

\section*{APPENDIX B}

We consider the random variable $\mathit{X=U^{2}+V^{2}},$ where the normalized variable $\mathit{U^{^{2}}/\sigma_{U}^{2}}$
follows a non-central chi-squared distribution, and $V^{2}$ is Gamma
distributed with shape parameter 1/2 and scale parameter 2$\sigma_{V}^{2}$. Consequently, $\left|H\right|^{2}$
can be expressed as the sum of a scaled non-central chi-squared variable and a Gamma variable. To characterize the distribution of $X$ we utilize its cumulative generating function, defined as K$_{X}(t)$ = ln$\mathbb{E}\left[e^{tX}\right]$. Since
$U$ and $V$ are independent (APPENDIX A)

\[
K_{X}(t)=K_{U^{2}}(t)+K_{V^{2}}(t),
\]
\begin{equation}
=\frac{\mu^{2}t}{1-2\sigma_{U}^{2}t}-\frac{1}{2}\ln(1-2\sigma_{U}^{2}t)-\frac{1}{2}\ln(1-2\sigma_{V}^{2}t).
\end{equation}

We now approximate for large $M$ the first term in the r.h.s. We
develop its Maclaurin series expansion as

\[
\frac{\mu^{2}t}{1-2\sigma_{U}^{2}t}=\mu^{2}t+\mu^{2}(2\sigma_{U}^{2})t^{2}+\mu^{2}(2\sigma_{U}^{2})t^{3}+....
\]
\[
=\frac{\mu^{2}}{4\sigma_{U}^{2}}\sum_{K=1}^{\infty}\frac{k}{2^{k-1}}\frac{(4\sigma_{U}^{2}t)^{k}}{k}
\]
\[
=\frac{\mu^{2}}{4\sigma_{U}^{2}}\sum_{k=1}^{\infty}\frac{(4\sigma_{U}^{2}t)^{k}}{k}-g(t)\sum_{k=1}^{\infty}\frac{(4\sigma_{U}^{2}t)^{k}}{k}
\]
\begin{equation}
=-\ln(1-4\sigma_{U}^{2}t),
\end{equation}

where

\[
g(t)=\frac{\mu^{2}}{4\sigma_{U}^{2}}\sum_{k=3}^{\infty}(1-\frac{k}{2^{k-1}})\frac{(4\sigma_{U}^{2}t)^{k}}{k}<\frac{\mu^{2}}{4\sigma_{U}^{2}}\sum_{K=3}^{\infty}\frac{(4\sigma_{U}^{2}t)^{k}}{k}
\]

\[
=4\mu^{2}\sigma_{U}^{2}t^{2}\sum_{k=1}^{\infty}\frac{(4\sigma_{U}^{2}t)^{k}}{k+2}<4\mu^{2}\sigma_{U}^{2}t^{2}\sum_{k=1}^{\infty}\frac{(4\sigma_{U}^{2}t)^{k}}{k}
\]
\begin{equation}
=-4\mu^{2}\sigma_{U}^{2}t^{2}\ln(1-4\sigma_{U}^{2}t).
\end{equation}

Given that $\sigma_{U}^{2}$=$O(M^{-1}),$ we have g(t)=$O(M^{-2}),$
such that $\left(45\right)$ is well approximated by 

\begin{equation}
K_{X}(t)=-\frac{\mu^{2}}{4\sigma_{U}^{2}}\ln(1-4\sigma_{U}^{2}t)-\frac{1}{2}\ln(1-2\sigma_{U}^{2}t)-\frac{1}{2}\ln(1-2\sigma_{V}^{2}t).
\end{equation}

The expression $\left(48\right)$ corresponds to the cumulative generating
function of the sum of three independent Gamma variables with (shape,
scale) parameters ($\frac{\mu^{2}}{4\sigma_{U}^{2}},4\sigma_{U}^{2}$),
(1/2 , 2$\sigma_{U}^{2})$, and (1/2 , 2$\sigma_{V}^{2})$, respectively.
While the resulting distribution can be characterized based on \cite{re044},
we simplify by making a further approximation

\begin{equation}
K_{X}(t)=-\frac{\mu^{2}}{4\sigma_{U}^{2}}\ln(1-4\sigma_{U}^{2}t),
\end{equation}

\noindent which has $O(M^{-1})$ error. Thus, for large $M,X$ has
a Gamma distribution with shape $\frac{\mu^{2}}{4\sigma_{U}^{2}}$
and scale $4\sigma_{U}^{2}$. It follows through variable transformation
that $\sqrt{X}$ has a Nakagami distribution with fading parameter
(shape) and spread $\mu^{2}$. Therefore,

\begin{equation}
m=-\frac{\mu^{2}}{4\sigma_{U}^{2}}=\frac{M}{2}\frac{\varphi_{1}^{2}\alpha^{4}}{1+\varphi_{2}-2\varphi_{1}^{2}\alpha^{4}}.
\end{equation}

\section*{Acknowledgment}

This work was supported and funded by the Deanship of Scientific Research at Imam Mohammad Ibn Saud Islamic University (IMSIU) (grant number: IMSIU-DDRSP2504).

\selectlanguage{english}%
\bibliographystyle{IEEEtran}
\bibliography{bib}
\selectlanguage{american}%

\end{document}